\documentclass[shortnote,twocolumn]{jpsj2}
\usepackage{graphicx}

\def\runtitle{Quasi-Particle density of states and Thouless conductance of
  disordered $d$-wave superconductors}

\def\runauthor{Bodo \textsc{Huckestein} and Alexander \textsc{Altland}}

\title{Quasi-Particle density of states and Thouless conductance of
  disordered $d$-wave superconductors}

\author{Bodo \textsc{Huckestein}$^1$ and Alexander \textsc{Altland}$^2$}
\inst{$^1$Institut f\"ur theoretische Physik III,
  Ruhr-Universit\"at Bochum, D-44780 Bochum, Germany\\
  $^2$Institut f\"ur theoretische Physik, Universit\"at zu
  K\"oln, D-50937 K\"oln, Germany}

\recdate{\today}

\abst{}

\kword{%
d-wave superconductor, disorder, density of states, localization,
Thouless numbers
}

\begin{document}
\sloppy

\maketitle

Disorder in electronic systems suppresses diffusion and can lead to
localization \cite{And58}. The origin of this behavior is enhanced
backscattering due to quantum interference. In addition, disorder
leads to a broadening of the density of states (DoS). However, the
localization properties of a disordered system are usually not
discernible from a study of the DoS. It is necessary to calculate
two-particle properties or consider the influence of boundary
conditions on energy eigenvalues as in the study of Thouless numbers
to obtain information about localization. Another feature of
disordered systems is the independence of their localization
properties from details of the disorder in the system. For example,
the range of the disorder affects non-universal quantities like the
mean-free path but does not influence whether or not the system shows
localization at all.

Recently, new symmetry classes came into focus that do not conform to
the standard expectations about disordered systems outlined above
\cite{AZ97}. As we will show in this contribution, details of disorder
do matter for these systems and localization properties leave their
mark on the DoS. In particular, we look at a system that presents an
idealization of a disordered $d$-wave superconductor. In our study we
neglect effects of self-consistency and concentrate on the $d$-wave
symmetry and influence of different kinds of disorder.

We consider the lattice quasi-particle Hamiltonian
\begin{equation}
  \label{eq:1}
  H = \sum_{ ij;\sigma} (t_{ij}-\mu\delta_{ij})
  c_{i\sigma}^\dagger c_{j\sigma}^{\phantom{\dagger}}
  + \sum_{ ij} \Delta_{ij}
  c_{i\uparrow}^\dagger c_{j\downarrow}^\dagger + \mathrm{h.c.},
\end{equation}
with the hopping matrix elements $t_{ij}$, chemical potential $\mu$,
and order parameter $\Delta_{ij}$. The sums run over points of a
two-dimensional square lattice with unit spacing and the operators
$c_{i\sigma}^\dagger$ create a spin-1/2 particle of spin $\sigma$ at
site $i$. In the following, we take only into account on-site
potentials and nearest-neighbor hopping, $t_{ij}=\epsilon_i\delta_{ij}
+ t\delta_{i,j\pm e_k}$, where $e_k$ is the unit vector in
$k$-direction. For convenience we set $\mu=0$ (the half-filled band.)
The order parameter $\Delta_{ij}=\Delta(\delta_{i,j\pm
  e_x}-\delta_{i,j\pm e_y})$ has $d_{x^2-y^2}$-symmetry. We study
random on-site potential $\epsilon_i$ with a spatial correlation
length $\xi$ \cite{HA00ASZ02}. The spectrum of (\ref{eq:1}) in the absence
of disorder contains four low-energy sectors located near the
wave-vectors $k=(\pm\pi/2,\pm\pi/2)$ (Dirac cones). Disorder couples
the states of the unperturbed Hamiltonian but for long-ranged
potentials ($\xi\gg1$) the coupling between the different Dirac cones
will be very small and will only be relevant on very large length
scales.

\begin{figure}[tbp]
  \begin{center}
    \includegraphics[width=7.6cm]{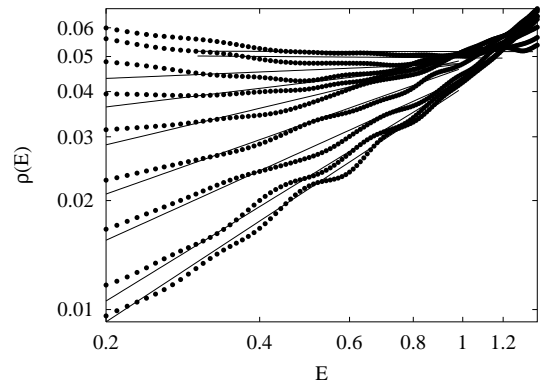}
    \caption{Double logarithmic plot of the density of states for
      long-ranged disorder ($\xi=2$). Disorder values are
      $W=1,2,\ldots,7,8,10$ (bottom to top). Dots 
      ($\bullet$) represent data and lines power law fits to the
      respective intervals.}
    \label{fig:2}
  \end{center}
\end{figure}

\begin{figure}[tbp]
  \begin{center}
    \includegraphics[width=7.6cm]{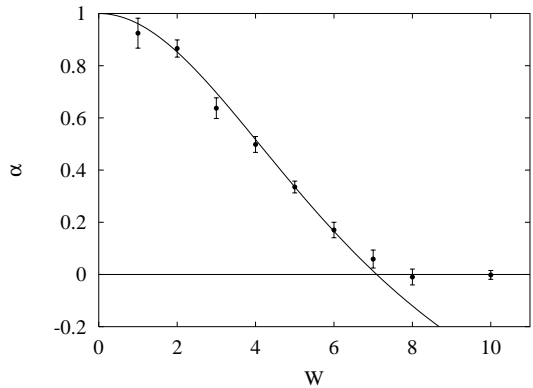}
    \caption{Exponents $\alpha$ extracted from the fitted curves in
      Fig.~\ref{fig:2} as a function of disorder $W$ for
      $\Delta=1$. The solid curve is the result eq.~(\ref{eq:6}).}
    \label{fig:4}
  \end{center}
\end{figure}

It has been predicted that for long-ranged disorder the spectrum of
eq.~(\ref{eq:1}) shows power law behavior with a non-universal
exponent corresponding to a line of critical states \cite{ludwig94,NTW94}:
\begin{equation}
  \label{eq:6}
  \rho(E) \sim |E|^\alpha, \qquad \alpha =
  \frac{1-g}{1+g}, \qquad g=\frac{W^2}{16\pi t\Delta}.
\end{equation}
We indeed find numerically power law behavior for intermediate
energies (Fig.~\ref{fig:2}) with the exponents well described by
eq.~(\ref{eq:6}) (Fig.~\ref{fig:4}). For larger values of the
disorder, $g>1$, the DoS does not diverge but remains constant as a
function of energy as has been argued by Gurarie \cite{gurarie}. In
the nomenclature of Altland and Zirnbauer the case of long-ranged
disorder conforms to the symmetry class $A$III \cite{AZ97}.

\begin{figure}[b]
  \begin{center}
    \includegraphics[width=7.6cm]{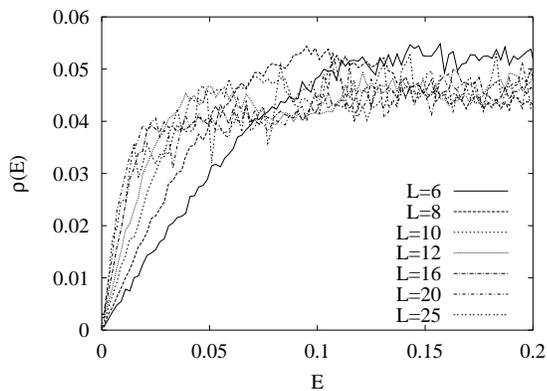}
    \caption{Density of states for $d=0.1$, disorder strength $W=9$,
      gap $\Delta=1$, and various system sizes $L$.}
    \label{fig:w9}
  \end{center}
\end{figure}

We now turn our attention to the case of short-ranged disorder,
belonging to the symmetry class $C$I. Here the DoS does not show power
law scaling and is only weakly energy dependent for strong disorder.
However, since the spectrum is particle-hole symmetric level repulsion
of the eigenvalues near zero energy creates a microgap with a linear
DoS, $\rho(E)\propto |E|$. The width of the microgap is given by the
mean level spacing of a single localization volume
$\Delta_\xi=(\rho_0\xi^2)^{-1}$ if the localization length $\xi$ is
smaller than the system size $L$. Otherwise, it is given by the mean
level spacing of the whole system. As a result, the width of the
microgap first shrinks with increasing system size and finally
saturates (see Fig.~\ref{fig:w9}).

\begin{figure}[tb]
  \begin{center}
    \includegraphics[width=7.6cm]{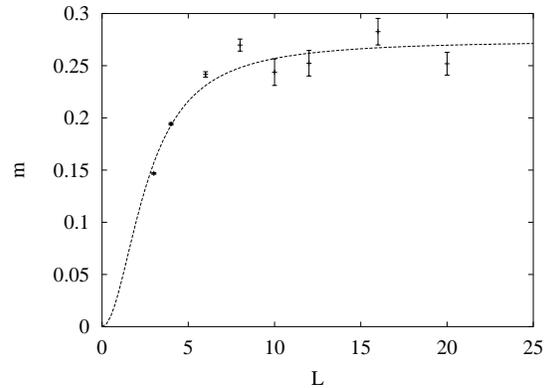}
    \caption{Slope $m$ of the microgap for correlation length $d=0.1$,
      disorder strength $W=18$, and gap $\Delta=1$ as a function of
      system size $L$. The dashed line is a fit of the function in
      Eq.~(\ref{eq:2}) with $m_\infty=0.27$ and $\xi=2.6$.}
    \label{fig:fit_12}
  \end{center}
\end{figure}

In an attempt to quantitatively analyze the crossover of the microgap
width from mean level spacing of the whole system to that of a single
localization volume, we have fitted the system size dependence of the
slope $m$ of the microgap DoS to the form
\begin{equation}
  \label{eq:2}
  m(L) = m_\infty \frac{\left(L/\xi\right)^2}{1+\left(L/\xi\right)^2}.
\end{equation}
For small system sizes this expression is consistent with the system
size dependence of the mean level spacing $\Delta\propto L^{-2}$ while
it approaches a disorder dependent constant at large
$L$. Fig.~\ref{fig:fit_12} shows that eq.~(\ref{eq:2}) indeed provides
a reasonable description of the crossover and that the localization
length can indeed be deduced from the DoS.

\begin{figure}[tb]
  \begin{center}
    \includegraphics[width=7.6cm]{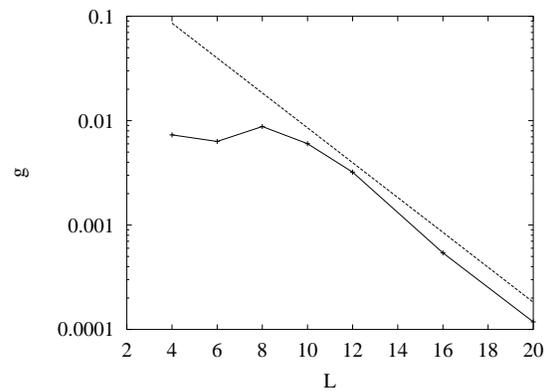}
    \caption{Semi-logarithmic plot of the Thouless numbers $g$ near
      zero energy for correlation length $d=0.1$, disorder strength
      $W=18$ and gap $\Delta=1$ as a function of system size $L$. The
      dashed line corresponds to an exponential decrease with
      localization length $\xi=2.6$.}
    \label{fig:thouless_18_log}
  \end{center}
\end{figure}

Finally, we can check whether the crossover length extracted from the
DoS corresponds to the localization length that governs transport. To
this end we have averaged the typical Thouless numbers $g$ over a
small energy interval near zero energy ($0<E<0.05$, less than a third of
the width of the microgap). Fig.~\ref{fig:thouless_18_log} compares
the system size dependence of this quantity with an exponential decay
with the crossover length obtained from Fig.~\ref{fig:fit_12} showing
good agreement.

To summarize, we have presented evidence that in disordered $d$-wave
superconductors details of disorder are relevant. For long-ranged
disorder, we find the expected power law behavior of the quasi-particle
density of states. For short-ranged disorder, we have shown that DoS
contains information about the localization properties of the system
in contrast to conventional disordered system. In particular, we could
extract the localization length from the system size dependence of the
microgap in the DoS. Calculations of the Thouless conductance of the
system corroborate our interpretation.

\end{document}